\documentclass[twocolumn,aps,floatfix,showpacs]{revtex4}
\usepackage{amsmath,amssymb,eucal,graphicx}

\begin{document}
\title{A ``Burnt Bridge'' Brownian Ratchet} 
\author{T. Antal}
\altaffiliation{On leave from Institute for Theoretical Physics -- HAS,
  E\"otv\"os University, Budapest, Hungary} 
\author{P. L. Krapivsky} \affiliation{Department of Physics, Boston
  University, Boston, Massachusetts 02215, USA}

\begin{abstract}
  
  Motivated by a biased diffusion of molecular motors with the bias
  dependent on the state of the substrate, we investigate a random walk on a
  one-dimensional lattice that contains weak links (called ``bridges'') which
  are affected by the walker. Namely, a bridge is destroyed with probability
  $p$ when the walker crosses it; the walker is not allowed to cross it again
  and this leads to a directed motion.  The velocity of the walker is
  determined analytically for equidistant bridges. The
  special case of $p=1$ is more tractable --- both the velocity and the
  diffusion constant are calculated for uncorrelated
  locations of bridges, including periodic and random distributions.
  
\end{abstract}

\pacs{02.50.-r, 05.40.Fb, 87.16.Nn} 
\maketitle

\section{Introduction}

The motion of a particle depends on the medium. Often the inverse is also
true, that is, the particle motion changes the medium. Such problems are
characterized by an infinite memory --- not only the present position of the
particle, but the entire past determines the future --- and they are usually
extremely difficult.  Perhaps the most famous example is the self-avoiding walk
which is a random walk on a lattice with the restriction that hops
to already visited sites are forbidden \cite{saw}.  Similarly in a
path-avoiding walk, a random walker is not allowed to go over already visited
links.  Here we examine a generalization of the path-avoiding walk in which
the medium is a lattice with two kinds of links, strong and weak: strong
links are unaffected by the walker while weak links, called bridges, ``burn''
when they are crossed by the walker. The random walker is not allowed to
cross a burnt bridge. Obviously the burnt bridge model reduces to the
path-avoiding walk if all links are weak.  We also investigate a stochastic
burnt bridge model \cite{mai} in which the crossing of an intact bridge
leads to burning only with a certain probability $p$, while with probability
$1-p$ the bridge remains intact.

The stochastic burnt bridge model is a simplification of models proposed to
mimic classical molecular motors \cite{fox,siam,der,lipo} with energy coming from ATP
hydrolysis \cite{motors}. Recently it has been shown \cite{sci} that the
stochastic burnt bridge model with two weakly coupled tracks (the walker
moves on the ladder and the hopping rate between the tracks is small compared
to the hopping rate along the tracks) accurately describes experimental
results on the motion of an activated collagenase on the collagen fibril.

We shall focus on the one track stochastic burnt bridge model. Forbidding the
crossing of burnt bridges essentially imposes a bias, and the goal is to
compute the velocity $v(c,p)$ and the diffusion coefficient $D(c,p)$ as
functions of the density of bridges $c$ and the bridge burning probability
$p$. The velocity and the diffusion coefficient also depend on the
positioning of the bridges. We shall tacitly assume that the bridges are
placed without correlations, and we shall often specify our findings to two
particularly interesting and natural positioning of the bridges --- a regular
equidistant spacing and a random distribution.

The rest of this paper is organized as follows. In the next section, we
describe various versions of the burnt bridge model and outline the major
results. Section \ref{sec:bbm} is devoted to the derivation of the velocity and the
diffusion coefficient for the burnt bridge model and a modified burnt bridge
model. In Sec.~\ref{SBB}, the stochastic burnt bridge model is studied, and the
velocity is computed for equidistant bridges.  Finally,
a few open questions are discussed (Sec.~\ref{disco}). Various calculations are
relegated to the Appendices.

\section{Main Results}
\label{results}

In all versions of a burnt bridge model, the walker asymptotically behaves as
a biased random walk. Mathematically this means that the probability of
finding the walker at position $x$ is a Gaussian centered around $\langle x
\rangle = vt$ with width $\langle x^2 \rangle-\langle x \rangle^2 = 2Dt$.

\begin{figure}[htb]
  \includegraphics[width=0.9\linewidth]{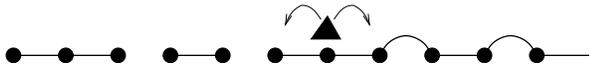}
  \caption{The random walker (filled triangle) hops to adjacent sites when it
    is away from the bridges. Strong links are shown by horizontal lines,
    intact bridges are depicted by arcs, and the absence of a link indicates
    a burnt bridge.}
  \label{demo}
\end{figure}

Away from the bridges, the walker hops to adjacent sites equiprobably
(Fig.~\ref{demo}). Thus if there were no bridges, the velocity would be equal
to zero and the diffusion constant would be equal to 1/2 (the lattice spacing
and the time step between the hops are set to unity). Bridges (which are
assumed to be intact initially) generate a directed motion.  The velocity
$v$ of the walker depends on the density of bridges $c$ and on the bridge
distribution.  Particularly simple expressions are obtained for periodically
and randomly positioned bridges
\begin{equation}
  \label{v}
  v(c)=
  \begin{cases}
    c        & {\rm periodic}\\
    c/(2-c) & {\rm random}
  \end{cases}
\end{equation}
Of course, in the periodic case the density $c$ attains only inverse integer
values ($1,1/2,1/3,\ldots$) while when bridges are placed at random the
density can attain any value $0<c\leq 1$.

In the realm of continuum approximation, the velocity $v(c)$ has been
computed by Mai {\em et al.}  \cite{mai}.  Unexpectedly, the continuum
approximation is exact in the periodic case; for random locations, the
continuum approximation gives $v(c)=c/2$. This result is asymptotically exact
in the small $c$ limit, albeit overall it is just an approximation which more
and more deviates from the exact result as the density approaches $c=1$
(that is, for the path-avoiding walk).

\begin{figure}[htb]
  \includegraphics[width=0.9\linewidth]{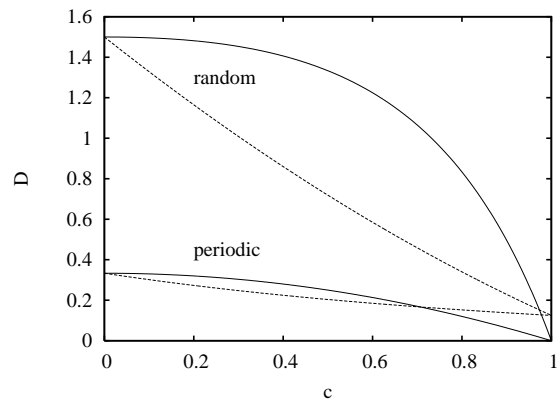}
  \caption{Diffusion coefficient in the original (solid) and in the modified
    (dashed) burnt bridge model, for random and regular bridge
    distributions.}
  \label{modD}
\end{figure}

The dependence of the diffusion coefficient on the density of bridges $c$ is
also relatively simple for periodically and randomly positioned bridges
\begin{equation}
  \label{Dc}
  D(c)=
  \begin{cases}
    \displaystyle{\frac{1}{3}(1-c^2)}  & {\rm periodic}\\\\
    \displaystyle{\frac{3}{2} \frac{1-c}{(1-c/2)^2}} & {\rm random}
  \end{cases}
\end{equation}
The diffusion coefficient monotonously decreases as $c$ increases (see
Fig.~\ref{modD}), and $D(1)=0$ since on a lattice fully covered by bridges,
the walker moves deterministically. The diminishing of $D(c)$ has apparently
been observed in Ref.~\cite{sci}.  Intriguingly, equation (\ref{Dc}) gives
$D_{\rm per}(+0)=1/3$ ($D_{\rm ran}(+0)=3/2$) which is smaller (larger) than
the ``bare'' diffusion coefficient $D_{\rm bare}=1/2$ that characterizes
diffusion on the one-dimensional lattice without bridges.  This sudden jump
of the diffusion coefficient occurs when the density becomes positive.  The
reason is that any positive $c$ (irrespectively however small it is) makes
the lasting influence on the fate of the walker who is forced to remain to
the right of the last burnt bridge. Thus the very rare burning events
substantially affect the diffusion coefficient.

\begin{figure}[htb]
  \includegraphics[width=0.9\linewidth]{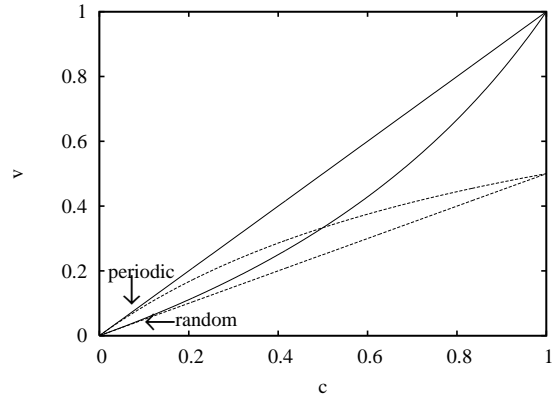}
  \caption{Velocity of the walker in the original (solid) and in the modified
    (dashed) burnt bridge model, for random and regular bridge
    distributions.}
  \label{modv}
\end{figure}

The details of the walker dynamics at the boundary of the burnt bridge affect
the velocity and the diffusion coefficient.  To illustrate this assertion
recall that in the framework of the burnt bridge model the walker at the
boundary of the burnt bridge always moves to the right.  Another natural
definition is to allow an attempt to cross the burnt bridge --- the attempt
fails and the walker remains at its position. We calculated the velocity for
this modified burnt bridge model \cite{again}:
\begin{equation}
  \label{v-refl}
  v(c)=
  \begin{cases}
    c/(1+c) & {\rm periodic}\\
    c/2 & {\rm random}
  \end{cases}
\end{equation}
We also computed the diffusion coefficient:
\begin{equation}
  \label{Dcmod}
  D(c)=
  \begin{cases}
    \displaystyle{\frac{1}{3}\,\frac{1}{1+c}-\frac{1}{6}\,\frac{c^2}{(1+c)^2}}
    & {\rm periodic}\\\\
    \displaystyle{\frac{3}{8}c^2-\frac{7}{4}c+\frac{3}{2}} & {\rm random}
  \end{cases}
\end{equation}
Perhaps the largest difference between the two models is that in the realm of
the modified burnt bridge model the walker never moves deterministically ---
even when $c=1$ it moves diffusively although the diffusion coefficient is
small, namely it is 4 times smaller than the bare diffusion coefficient. Not
surprisingly, the quantitative predictions of the two models are
substantially different when $c$ is large (see Figs.~\ref{modv} and \ref{modD}).

\begin{figure}[htb]
  \includegraphics[width=0.9\linewidth]{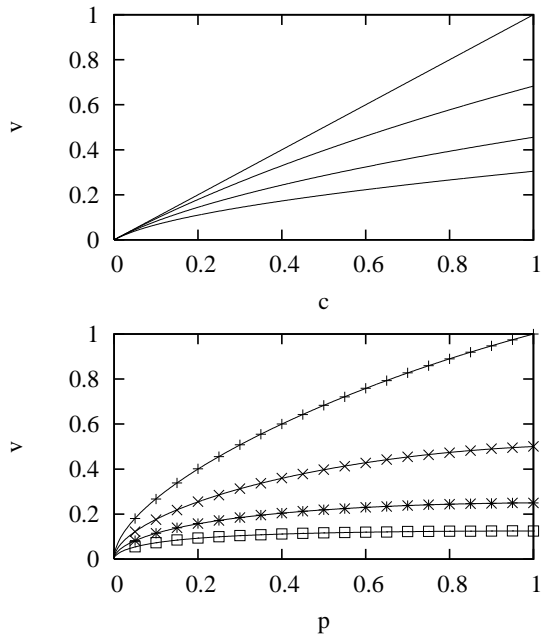}
  \caption{The velocity $v(c,p)$ vs.\ $c$ (for $p=1, 1/2, 1/4, 1/8$, top to
    bottom) and $p$ (for $c=1, 1/2, 1/4, 1/8$, top to bottom).  The lattice
    is periodically covered by bridges. The symbols are the results of
    numerical simulations.}
  \label{vel}
\end{figure}

For the stochastic burnt bridge model, we computed only the velocity, and only for
periodically located bridges.  We found
\begin{equation}
  \label{v-main}
  v(c,p)=\frac{cp}{cp+2-p}\cdot
  \frac{2-p+V}{p(1-c)+V}
\end{equation}
where we used the shorthand notation
\begin{equation*}
  V=\frac{p(2-p)(1-c)}{2}\left\{-1+
    \sqrt{1+\frac{4c}{p(2-p)(1-c)^2}}\right\}
\end{equation*}
For $p=1$, Eq.~(\ref{v-main}) agrees with already known result $v=c$ [see
Eq.~(\ref{v})]; for $c=1$ (the lattice fully covered by bridges), the
velocity is given by the following neat expression
\begin{equation}
  \label{vp-eq}
  v(1,p) = \frac{p+\sqrt{p(2-p)}}{2}
\end{equation}
{}From Eq.~(\ref{v-main})  (see also Fig.~\ref{vel}) one
finds the asymptotics
\begin{equation}
  \label{v-asymp}
  v(c,p)\to 
  \begin{cases}
    c            & {\rm when} ~~c\ll p\\
    \sqrt{cp/2} & {\rm when} ~~p\ll c
  \end{cases}
\end{equation}

For $c\ll p$, the distance between neighboring bridges is large. Thus the walker typically crosses the next bridge several times and hence almost all bridges get burnt.  
Therefore the $p=1$ results ought to be recovered. 
Equation (\ref{v-asymp}) shows that this is indeed correct in the
periodic case; in the random case (where we do not know an exact solution) we
similarly expect $v(c,p)\approx c/2$ when $c\ll p$. In the complimentary
limit $p\ll c$, the walker on average makes many steps before the burning
occurs, and it is intuitively obvious that we can renormalize $c\to 1$ and
simultaneously $p\to cp$. Hence $v(c,p)\to v(1,cp)$ when $p\ll c$, and
therefore the asymptotics given in (\ref{v-asymp}) can also be extracted from
the simple solution (\ref{vp-eq}).

\begin{figure}[htb]
   \includegraphics[width=0.9\linewidth]{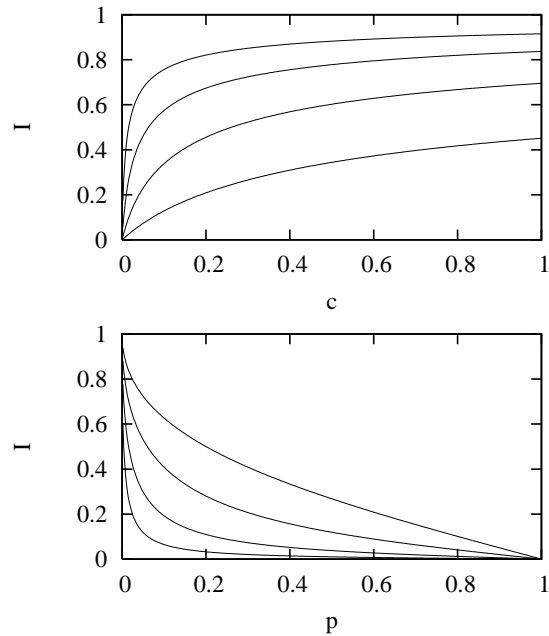}
   \caption{The fraction of intact bridges, $I(c,p)$ vs.\ $c$ (for $p=1/4,
     1/16, 1/64, 1/256$, from bottom to top), and vs.\ $p$ (for $p=1, 1/4,
     1/16, 1/64$, from top to bottom) in the periodic case.}
   \label{intact}
\end{figure}

Another interesting and experimentally accessible quantity is the fraction of
bridges left intact by the walker. In the long time limit it can be expressed
as
\begin{equation}
 I(c,p) =  1 - \frac{V}{c}\cdot\frac{cp+2-p}{V+2-p}
\label{intactic}
\end{equation}
Figure \ref{intact} shows that the fraction of intact bridges is a decreasing
function of $p$ for fixed $c$ (this is intuitively obvious) and an increasing
function of $c$ for fixed $p$. This latter feature is understood by noting
that on average a bridge is visited more often when the density of bridges
gets smaller.

\section{The burnt bridge model}
\label{sec:bbm}

The walker hops $x\to x\pm 1$ equiprobably when it is away from the burnt
bridges.  The lattice spacing and the time step between successive hops are
set to unity, and therefore the diffusion coefficient is $D=1/2$ for the
lattice with no bridges.  We assume that initially all bridges are intact.
Let the walker cross a bridge for the first time from the left. This implies
that the walker will never cross it again and determines the fate of the
walker, namely the walker will drift to the right and the closest burnt
bridge will always be on the left. By definition, when the walker is on the
right boundary of the burnt bridge, it always makes the step to the right.

Two special positioning of the bridges are regular (equidistant locations)
and random. These are two most interesting cases, although the formalism can
be straightforwardly extended to arbitrary distributions as long as the
distances between neighboring bridges are uncorrelated.

\subsection{Velocity and Diffusion coefficient}
\label{vD}

The walker successively crosses the intact bridges which immediately turn
into burnt bridges. Instead of the original walker it proves convenient to
consider an equivalent walk whose position at time $t$ is on the right edge
of the the last burnt bridge. This walk is still discrete in space and time,
but the step length is now equal to the distance between successive bridges,
and the time between the steps is at least as large as the step length.  In
general, we denote by $\Psi(x,t)$ the probability that the walk makes a step
of length $x$ after waiting $t$ units of time.

The following derivation is essentially the discrete time version of the
so-called continuous time random walk (CTRW) \cite{{ctrw},{Hughes}}.  This
title refers to the continuous nature of the the waiting time distribution,
while in our model the discrete nature of time is essential, and therefore we
present the complete derivation of the necessary results.  Note also that
even though our walk always steps to the right, the derivation is the same
as for a general walk which can step in both directions; hence, we present a
general derivation. 

Let $\Psi_j(x,t)$ be the probability that the walk arrives at site $x$ at
time $t$ and at the $j$-th step, and hence $\Psi_1(x,t)\equiv\Psi(x,t)$. This
probability satisfies the recurrence formula
\begin{equation}
\label{conv} 
\Psi_j(x,t) = \sum_{x'=-\infty}^\infty \sum_{t'=0}^t  \Psi_{j-1}(x-x',t-t') \Psi(x,t) ~.
\end{equation}
Using the generating function (the discrete version of the Fourier-Laplace
transform) 
\begin{equation}
  \Psi_j(q,u) = \sum_{x=-\infty}^\infty \sum_{t=0}^\infty \Psi_j(x,t) q^x u^t 
\end{equation}
we transform the convolution (\ref{conv}) into the product
\begin{equation}
  \Psi_j(q,u) = \Psi_{j-1}(q,u) \Psi(q,u) =  [\Psi(q,u)]^j ~.
\end{equation}
The probability $Q(x,t)$ of arriving at site $x$ at time $t$ (after an arbitrary
number of steps) is
\begin{equation}
  Q(x,t) = \sum_{j=0}^\infty \Psi_j(x,t)
\end{equation}
The corresponding generating function can be written in a closed form
\begin{equation}
  Q(q,u) = \sum_{j=0}^\infty [\Psi(q,u)]^j = \frac{1}{1-\Psi(q,u)} ~.
  \label{Qfin}
\end{equation}

The probability $P(x,t)$ that the walk is at site $x$ at time $t$ can be
obtained by noting that in order to be at site $x$ at time $t$, the walk has
to arrive at site $x$ not later than at time $t$ and has to stay there until
$t$. Hence
\begin{equation}
  P(x,t) = \sum_{t'=0}^t \phi(t-t')Q(x,t')
  \label{npq}
\end{equation}
where $\phi(t)$ is the probability that the walk does not move for a time
interval $t$
\begin{equation}
  \phi(t) = 1 -  \sum_{t'=0}^t \Psi(t')
  \label{phit}
\end{equation}
The probability $\Psi(t)$ that the walk makes at least one step during the
time interval $t$ is obtained by summing over all possible step lengths
\begin{equation}
  \Psi(t) = \sum_{x=-\infty}^\infty \Psi(x,t) ~.
\label{Psitdef}
\end{equation}
Using Eq.~(\ref{phit}) we compute the generating function of $\phi(t)$
\begin{equation}
  \phi(u) = \frac{1-\Psi(u)}{1-u}
\end{equation}
The generating function of $P(x,t)$ is now easily derived since
Eq.~(\ref{npq}) is a convolution:
\begin{equation}
  P(q,u) = \phi(u) Q(q,u) = \frac{1-\Psi(u)}{(1-u)[1-\Psi(q,u)]}
  \label{genfinal}
\end{equation}

We are going show that position of the walk follows a Gaussian probability
distribution for large times
\begin{equation}
  \label{Gaussian}
  P(x,t) = \frac{1}{\sqrt{4\pi Dt}}\,
  \exp\left\{-\frac{(x - vt)^2}{4Dt}\right\}
\end{equation}
which is centered around a mean value $\langle x\rangle=vt$ with a mean
square deviation $\langle x^2\rangle - \langle x\rangle^2 = 2Dt$, with $v$
being the speed of the walk, and $D$ being the diffusion coefficient. We
define the new variables $\gamma$ and $\epsilon$ as $q=e^{i\gamma}$ and
$u=e^{-\epsilon}$. In order to show that the long time limit of $P(x,t)$
is Gaussian, we ought to show that in the $\gamma,
\epsilon\to0$ limit $P(\gamma, \epsilon)$ is equal to the Laplace-Fourier transform of the Gaussian
(\ref{Gaussian})
\begin{equation}
  P(\gamma, \epsilon) = \frac{1}{\epsilon - i\gamma v + \gamma^2D} ~.
  \label{Gausslap}
\end{equation}
Therefore we must show that $P(\gamma, \epsilon)$
given by Eq.~(\ref{genfinal}) attains the form of Eq.~(\ref{Gausslap}).

Although the probability $\Psi(x,t)$ is not separable in general, it can
always be written as the product 
\begin{equation}
  \Psi(x,t) =  S(x)\,\Psi(t|x) 
\label{cond}
\end{equation}
of the probability $S(x)$ that the next step has length
$x$ (distance to the next bridge) times the conditional probability
$\Psi(t|x)$ that the next step happens after $t$ waiting time, given that the
length of this step is $x$.

Now we calculate $\Psi(\gamma, \epsilon)$ up to the second order in $\gamma$
and $\epsilon$.  First, we calculate the generating function with respect to
time in the $\epsilon\to 0$ limit. Plugging (\ref{cond}) into 
\begin{equation*}
  \Psi(x,\epsilon) = \sum_{t\geq 0}\Psi(x,t)\,e^{-\epsilon t}
\end{equation*}
and expanding in $\epsilon$  up to the second order we obtain
\begin{equation*}
  \Psi(x,\epsilon) = S(x) \left( 1-\epsilon[t]_x + \frac{\epsilon^2}{2} [t^2]_x \right) 
\end{equation*}
where the moments of time are calculated at some fixed $x$ length of interval
\begin{equation}
  [ t^n ]_x = \sum_{t=0}^\infty t^n \Psi(t|x) ~.
  \label{momt}
\end{equation}
Performing the Fourier transform of $\Psi(x,u)$ and taking the $\gamma\to 0$
limit, we arrive at
\begin{equation*}
  \Psi(\gamma, \epsilon) = 1 - \epsilon\langle [t]_x \rangle + i\gamma\langle x \rangle
  -i \epsilon\gamma\langle x[t]_x \rangle + \frac{\epsilon^2}{2}\langle [t^2]_x \rangle
  - \frac{\gamma^2}{2}\langle x^2 \rangle
\end{equation*}
up to second order in $\gamma$ and $\epsilon$.  We also need to calculate
$\Psi(u)$, the generating function of $\Psi(t)$ defined by
Eq.~(\ref{Psitdef}). It is sufficient to know it only up to first order in
$\epsilon$:
\begin{equation*}
  \Psi(\epsilon) = 1 - \epsilon\langle [t]_x \rangle 
\end{equation*}

Plugging the above results into Eq.~(\ref{genfinal}) we find that up to first
order in both $\gamma$ and $\epsilon$, the quantity $\Psi(\gamma, \epsilon)$
attains the form $P(\gamma, \epsilon) = (\epsilon - i\gamma v)^{-1}$
with 
\begin{equation}
  v = \frac{\langle x \rangle}{\langle [t]_x \rangle} ~.
\label{vgen}
\end{equation}
Thus $P(x,t)$ is centered around $\langle x\rangle=vt$ with velocity given by
Eq.~(\ref{vgen}).

Calculating $P(\gamma, \epsilon)$ up to second order, and using the first
order expression for $\epsilon\approx-i\gamma v$ in the terms
containing $\epsilon \gamma$ and $\epsilon^2$, we arrive at
Eq.~(\ref{Gausslap}) with the diffusion coefficient given by
\begin{equation}
  D = \frac{\langle x^2 \rangle}{2\langle [t]_x \rangle} 
  + \frac{\langle [t^2]_x \rangle\langle x \rangle^2}{2\langle [t]_x \rangle^3} 
  - \frac{\langle x[t]_x \rangle\langle x \rangle}{\langle [t]_x \rangle^2} ~.
  \label{Dgen}
\end{equation}
Thus we conclude that $P(x,t)$ is indeed Gaussian.  Note that this result 
applies to any random walk --- discrete or continuous ---
where the steps are uncorrelated and all of the moments used in Eq.~(\ref{Dgen}) are finite.
Specifically, it applies to the burnt bridge model ($p=1$) if the distances
between bridges are uncorrelated.

\subsubsection{Special cases}

Consider first equidistant bridges separated by distance $\ell$. 
Then $S(x)=\delta_{x,\ell}$, and therefore $\langle
x\rangle=\ell,\langle x^2 \rangle=\ell^2$, etc.  The velocity (\ref{vgen})
and diffusion coefficient (\ref{Dgen}) simplify to
\begin{equation*}
v=\frac{\ell}{[t]_\ell}~,\quad  
D = \frac{\ell^2}{2} \frac{ [t^2]_\ell- [ t ]_\ell^2}{[ t ]_\ell^3}
\end{equation*}
Using the moments of time computed in Appendix~\ref{moments}
[Eqs.~(\ref{TL}) and (\ref{TL-per})], we arrive at Eqs.~(\ref{v}) and
(\ref{Dc}).

For randomly distributed bridges, the probability that the walk makes a step
of length $x$, that is the probability of having two neighboring bridges at
distance $x>0$, is
\begin{equation}
  S(x) = c(1-c)^{x-1} 
\end{equation}
We again use Eqs.~(\ref{TL}) and (\ref{TL-per}) for the
moments of time, and we also need the first four moments of $x$
\begin{equation}
  \begin{split}
    &\langle x \rangle = \frac{1}{c}\\
    &\langle x^2 \rangle = \frac{2-c}{c^2}\\
    &\langle x^3 \rangle = \frac{6-6c+c^2}{c^3}\\
    &\langle x^4 \rangle = \frac{24-36c+14c^2-c^3}{c^4} ~.
  \end{split}
\end{equation}
Using these expressions in Eqs.~(\ref{vgen}) and (\ref{Dgen}), we obtain
the velocity and the diffusion coefficient given by Eqs.~(\ref{v}) and
(\ref{Dc}), respectively.

Finally, consider the bimodal distribution 
\begin{equation}
\label{bimodal} 
  S(x) = q_1\,\delta_{x,\ell_1}+q_2\,\delta_{x,\ell_2}
\end{equation}
with two possible separations between the bridges, $\ell_1$ and $\ell_2$,
occurring independently with respective probabilities $q_1$ and $q_2$
(of course, $q_1, q_2\geq 0$ and $q_1+q_2=1$).   In this situation, the
velocity (\ref{vgen}) becomes 
\begin{equation*}
v=\frac{q_1\ell_1+q_2\ell_2}{q_1\ell_1^2+q_2\ell_2^2} 
\end{equation*}
and the diffusion coefficient (\ref{Dgen}) turns into
\begin{eqnarray}
\label{D-bimodal} 
D&=&\frac{1}{2}
-\frac{(q_1\ell_1^3+q_2\ell_2^3)(q_1\ell_1+q_2\ell_2)}{(q_1\ell_1^2+q_2\ell_2^2)^2}
-\frac{1}{3}\,\frac{(q_1\ell_1+q_2\ell_2)^2}{(q_1\ell_1^2+q_2\ell_2^2)^2}\nonumber\\
&+&\frac{5}{6}\,\frac{(q_1\ell_1^4+q_2\ell_2^4)(q_1\ell_1+q_2\ell_2)^2}
{6(q_1\ell_1^2+q_2\ell_2^2)^3}
\end{eqnarray}

For the system with bimodal bridge distribution (\ref{bimodal}), and
apparently for an arbitrary uncorrelated positioning of bridges, the
diffusion coefficient exceeds that of the corresponding periodic system at
the same bridge density. This general assertion is easy to verify in a
particularly interesting case when the density of bridges vanishes. Taking
the limit $\ell_1,\ell_2\to\infty$, and keeping the ratio 
$\ell_1/\ell_2=r (\leq 1)$ constant, we recast Eq.~(\ref{D-bimodal}) into
\begin{equation} 
\label{D-a} 
D=\frac{1}{2}+\left(\frac{q_1 r+q_2}{q_1 r^2+q_2}\right)^2
\left[\frac{5}{6}\,\frac{q_1 r^4+q_2}{q_1 r^2+q_2}
-\frac{q_1 r^3+q_2}{q_1 r+q_2}\right]
\end{equation}
A straightforward analysis of Eq.~(\ref{D-a}) shows that the diffusion
coefficient is larger than 1/3, which is the diffusion coefficient in the periodic case
($q_1=0$ or $q_2=0$). {}From Eq.~(\ref{D-a}) one finds that for $a\ll 1$, the
maximal diffusion coefficient, approximately $D\approx
\frac{10}{81}\,r^{-2}$, is achieved when the system is predominantly composed
of shorter segments $\ell_1$, namely when $q_2\approx r^2/2$. Therefore a
``superposition'' of two equidistant distributions, each characterized by the
diffusion coefficient 1/3, may have an arbitrarily large diffusion
coefficient.

\subsubsection{Simulations}

The velocity and the diffusion coefficient of the walker are determined using
the basic formulas $v=\langle x\rangle/t$ and $D=(\langle x^2\rangle -
\langle x\rangle^2)/2t$. Since the motion is self-averaging, simulating a
single walker for a long time is sufficient to obtain the velocity. To
measure the diffusion coefficient, however, one has to perform averages over
several runs. In the case of randomly distributed bridges, one also has to
average over the bridge distribution.

\begin{figure}[htb]
  \includegraphics[width=0.9\linewidth]{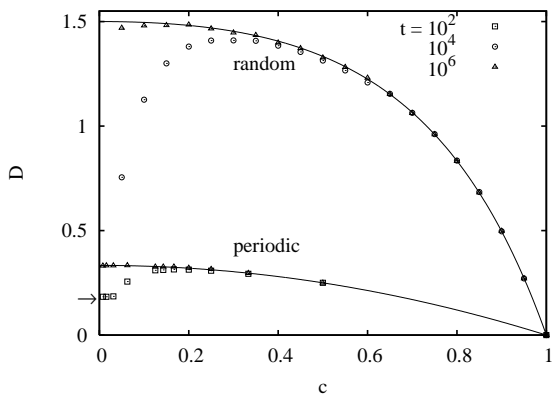}
  \caption{The diffusion coefficient $D(c)$ for the random and periodic
    bridge locations. Results of the simulations are also displayed for
    comparison. The arrow points to the theoretical value $D=1/2-1/\pi$ corresponding to a random walk 
    with a reflecting boundary.}
  \label{diff}
\end{figure}

Figure~\ref{diff} shows numerical results for the diffusion coefficient at
various times. The convergence to the theoretical predictions is slow when
the density of bridges is small. During a short time interval the walker does
not reach the second bridge, and actually behaves as a simple random walk
with a reflecting boundary at the origin.  Hence the probability of finding
the particle at position $x\ge0$ is a Gaussian centered around the origin,
and the formal definition of the diffusion coefficient yields $D=1/2-1/\pi$.
For the time intervals large compared to the time (of the order of $c^{-2}$)
between overtaking successive bridges, the coarse-grained motion becomes
similar to a biased random walk with the diffusion coefficient approaching
the theoretical predictions: $D(+0)=1/3$ in the periodic case and $D(+0)=3/2$
in the random case.

\subsection{Correlation function}

A correlation function measured experimentally in Ref.~\cite{sci} is
apparently proportional \cite{sci2} to the probability $C(t)$ that the walker
at the site $x_0$, will be at the same position time $t$ later.

As a warm-up, consider the extreme cases of the lattice without bridges
($c=0$) and the lattice fully covered by bridges ($c=1$). In the former case,
the correlation function obviously vanishes for odd $t$ while for even $t$ it
is given by the well-known expression
\begin{equation}
  \label{C0}
  C(2t)=2^{-2t} {2t \choose t}
\end{equation}
Note that the correlation function decays algebraically in the large time
limit:
\begin{equation}
  \label{C0-inf}
  C(2t)\to \frac{1}{\sqrt{\pi t}}\quad {\rm as} \quad t\to\infty
\end{equation}
For the lattice fully covered by bridges the walker can move only to the
right, the probability of not making a step is 1/2, and therefore the
correlation function
\begin{equation}
  \label{C1}
  C(t)=2^{-t} 
\end{equation}
is purely exponential.

In the general case $0<c<1$, the walker cannot leave the ``cage'' formed by
two neighboring bridges. As always, we consider the cage with sites
$x=0,\ldots,\ell-1$.  For simplicity, let's assume again that the initial
position is $x_0=0$.  Rather than considering the walker in the cage
$(0,\ell-1)$ with a special behavior at $x=0$ and the absorbing boundary at
$x=\ell$ one can analyze the ordinary random walker in the extended cage
$(-\ell,\ell)$ with absorbing boundaries at $x=\ell$ and $x=-\ell$. The
correlation function is merely the probability that this ordinary random
walker will be at $x=0$ at time $t$ and will remain inside the extended cage
in intermediate times.  This is a classic problem in probability theory
whose solution is a cumbersome sum of expressions like (\ref{C0}) with
alternating (positive and negative) signs. Therefore we employ a continuum
approximation which becomes asymptotically exact when $\ell\gg 1$ (and
accordingly $c=\ell^{-1}\ll 1$).  The solution is an infinite series of
exponentially decaying terms. Keeping only the dominant term we obtain
\begin{equation}
  \label{Ct}
  C(t)\to c\,\exp\left\{-\frac{\pi^2\,c^2}{8}\,t\right\}
\end{equation}
as $t\to\infty$. More precisely, the asymptotics (\ref{Ct}) is valid when
$t\gg c^{-2}$. In the regime $1\ll t\ll c^{-2}$, the dominant asymptotics is
the same as in the $c=0$ case, that is, $C(t)\sim t^{-1/2}$. It is therefore
understandable that a formula
\begin{equation}
  \label{fit}
  C(t)\approx (1+t)^{-1/2}\,\exp\left\{-\frac{\pi^2\,c^2}{8}\,t\right\}
\end{equation}
fits well experimental data (and indeed it does \cite{sci}). Yet the
true asymptotic behavior, Eq.~(\ref{Ct}), is purely exponential
without the power-law correction of Eq.~(\ref{fit}).

\subsection{Modified burnt bridge model}
\label{another}

The precise definition of the walker dynamics at the boundary of the burnt
bridge affects the results. We assumed that the walker at the boundary of the
burnt bridge always moves to the the right. Recall, however, that in the
stochastic version ($p<1$) when the walker attempts to hop over the bridge
from the left and the bridge burns, the walker actually remains at the same
position. This suggests to modify the rule at the boundary of the burnt bridge
 --- the walker either moves one step to the right or remains at the
same position if it has tried the forbidden move across the burnt bridge.
This defines the modified burnt bridge model.

The calculation of $v(c)$ and $D(c)$ goes along the same lines as for the
original burnt bridge model (Sec.~\ref{vD}) and leads to the
results presented in Sec.~\ref{results} and displayed on Figs.~\ref{modv} and
\ref{modD}.

For $c=1$, the diffusion coefficient of Eq.~(\ref{Dcmod}) is the same
$D(1)=1/8$ in both the periodic and the random case. This particular result also
follows from an independent calculation which we present here as it provides
a good check of self-consistency. The key simplifying feature of the lattice
fully covered with bridges is that the walker never hops to the left. The
position $x_t$ of the walker after $t$ time steps satisfies
\begin{equation}
  \label{xt}
  x_{t+1}=
  \begin{cases}
    x_t   & {\rm probability}~~ 1/2\\
    x_t+1 & {\rm probability}~~ 1/2
  \end{cases}
\end{equation}
from which
\begin{equation}
  \label{xt-av}
  \langle x_{t+1}\rangle=\langle x_t\rangle+\frac{1}{2}
\end{equation}
and
\begin{equation}
  \label{xt-av2}
  \langle x_{t+1}^2\rangle=\langle x_t^2\rangle+\langle x_t\rangle+\frac{1}{2}
\end{equation}
The variance $\sigma_t=\langle x_t^2\rangle-\langle x_t\rangle^2$ satisfies a
simple recurrence
\begin{equation}
  \label{st}
  \sigma_{t+1}=\sigma_t+\frac{1}{4}
\end{equation}
which follows from Eqs.~(\ref{xt-av})--(\ref{xt-av2}).  The initial condition
$x_0=0$ implies $\langle x_0\rangle=\sigma_0=0$.  Solving (\ref{xt-av}),
(\ref{st}) subject to these initial values we obtain
\begin{equation}
  \label{av}
  \langle x_t\rangle=\frac{1}{2}\,t~,~~
  \sigma_t=\frac{1}{4}\,t
\end{equation}
The velocity and the diffusion coefficient can be read off the general
relations $\langle x_t\rangle=vt$ and $\sigma_t=2Dt$.  Thus we recover the
already known value $v(1)=1/2$ and obtain the diffusion coefficient
$D(1)=1/8$ (which happens to be 4 times smaller than the bare diffusion
coefficient).

\section{The stochastic burnt bridge Model}
\label{SBB}

Apart from randomness in hopping, the stochastic burnt bridge model has an additional
stochastic element --- crossing the bridge leads to burning with probability
$p$ while with probability $1-p$ the bridge remains intact. To avoid the
possibility of trapping we additionally assume that if the particle attempts
to cross the bridge from the right and the bridge burns, the attempt is a
failure and the walker does not move. We have succeeded in computing $v(c,p)$
in the situation when the bridges are equidistant. We again employ an
approach involving auxiliary functions $T(x)$ (see Appendix \ref{moments})
and $L(x)$ (defined below). Perhaps, the entire problem can be treated by a
direct approach discussed in Appendix \ref{direct}, but that approach is more
lengthy and we have only succeeded in computing the velocity for $c=1$ that
way.

We must determine the average position of the first bridge that burns, and
the average time of that event. The walker starts at $x=0$, but it is again
useful to consider a more general situation when the walker starts at an
arbitrary position $x$. Denote by $L(x)$ the average position of the walker
at the moment when the first bridge burns. The walker hops $x\to x\pm 1$, and
therefore
\begin{equation}
  \label{Lx}
  L(x)=\frac{1}{2}\left[L(x-1)+L(x+1)\right]
\end{equation}
when $x\ne n\ell-1, n\ell$ with $n=1,2,3,\ldots$. On the boundaries of the
bridges the governing equation (\ref{Lx}) should be modified to account for
possible burning events:
\begin{subequations}
  \begin{align}
    &L(n\ell-1)=\frac{L(n\ell-2)+(1-p)L(n\ell)+pn\ell}{2}
    \label{n-1}\\
    &L(n\ell)=\frac{L(n\ell+1)+(1-p)L(n\ell-1)+pn\ell}{2}
    \label{n}
  \end{align}
\end{subequations}
Equation (\ref{Lx}) shows that $L(x)$ is a linear function of $x$ on each
interval between the neighboring bridges, i.e.,
\begin{equation}
  \label{Lx-lin}
  L(x) = A_n +(x-n\ell)B_n 
\end{equation}
for $n\ell\leq x\leq (n+1)\ell-1$. Plugging (\ref{Lx-lin}) into (\ref{n-1})
we obtain
\begin{equation}
  \label{An}
  A_{n-1} + \ell B_{n-1}=(1-p)A_n+pn\ell
\end{equation}
Similarly, equation (\ref{n}) reduces to
\begin{equation}
  \label{Bn}
  A_n=B_n+(1-p)[A_{n-1}+(\ell-1)B_{n-1}]+pn\ell
\end{equation}
Using (\ref{An}), we get rid of $B$'s in (\ref{Bn}) and obtain
\begin{equation}
  \label{ABn}
  A_{n-1} - 2gA_n + A_{n+1}=-\frac{p\ell}{1-p}-
  pn\ell\,\frac{p+(2-p)\ell}{1-p}
\end{equation}
Here we used a shorthand notation
\begin{equation}
  \label{g-def}
  g=\frac{p(2-p)(\ell-1)+2}{2(1-p)}
\end{equation}
The recurrence (\ref{ABn}) admits a general solution
\begin{equation}
  \label{An-sol}
  A_n=n\ell + \alpha + A_+\lambda_+^n + A_-\lambda_-^n
\end{equation}
where $A_n=n\ell + \alpha$ with $\alpha=\ell/[p+(2-p)\ell]$ is a particular
solution of the inhomogeneous equation (\ref{ABn}); the remaining
contribution $A_+\lambda_+^n + A_-\lambda_-^n$ with
\begin{equation}
  \label{lambda}
  \lambda_\pm=g\pm \sqrt{g^2-1}
\end{equation}
is the general solution of the homogeneous part of (\ref{ABn}).

If the walker is initially located far away from the origin, $x\gg \ell$, the
first bridge would burn somewhere in its proximity, that is $L(x)\sim x$.
This in conjunction with (\ref{Lx-lin}) lead to $A_n-n\ell=O(1)$ when $n\gg
1$.  On the other hand, the general solution (\ref{An-sol}) grows
exponentially since $\lambda_+>1$. This shows that the corresponding
amplitude must vanish: $A_+=0$. Since $L(x)$ is constant on the interval
$0\leq x\leq \ell-1$, we have $B_0=0$, or [see (\ref{An})]
\begin{equation}
  \label{A1}
  A_0=(1-p)A_1+p\ell
\end{equation}
By inserting $A_0=\alpha + A_-$ and $A_1=\ell+\alpha +A_-\lambda_-$ into
(\ref{A1}) and solving for $A_-$ we get
\begin{equation}
  \label{A-}
  A_-=\frac{\ell-p\alpha}{1-(1-p)\lambda_-}
\end{equation}

Return now to the situation when the walker starts at the origin. The average
displacement of the walker after the first burning event is $\langle
x\rangle=L(0)=A_0=\alpha+A_-$, or
\begin{equation}
  \label{xav-SBB}
  \langle x\rangle=\frac{\ell}{p+(2-p)\ell}
  \left[1+\frac{(2-p)\ell}{1-(1-p)\lambda_-}\right]
\end{equation}
In the limiting cases $p=1$ and $\ell=1$ we indeed recover $\langle
x\rangle=\ell$ and (\ref{avx-exact}), respectively.

The second part of the program is to determine the average time when the
first burning occurs. Again we choose to investigate a more general quantity
$T(x)$. It satisfies Eq.~(\ref{master}) when $x\ne n\ell-1, n\ell$. On the
boundaries of the bridges, the governing equations become
\begin{subequations}
  \begin{align}
    &T(n\ell-1)=\frac{T(n\ell-2)+(1-p)T(n\ell)}{2}+1
    \label{Tn-1}\\
    &T(n\ell)=\frac{T(n\ell+1)+(1-p)T(n\ell-1)}{2}+1
    \label{Tn}
  \end{align}
\end{subequations}

We seek a solution of (\ref{master}), (\ref{Tn-1}), (\ref{Tn}) which is
invariant under the transformation $x\leftrightarrow -x$, and periodic in the
large $x$ limit. A solution to equation (\ref{master}) is quadratic in $x$,
viz. $-x^2+Yx+Z$ with arbitrary $Y,Z$. The same holds in our situation except
that solutions in different intervals between the neighboring bridges differ.
Thus
\begin{equation}
  \label{Tx-quad}
  T(x)=-(x-n\ell)^2+(x-n\ell)Y_n +Z_n 
\end{equation}
Plugging (\ref{Tx-quad}) into (\ref{Tn-1})--(\ref{Tn}) we obtain
\begin{eqnarray*}
  \ell Y_{n-1}+Z_{n-1}&=&(1-p)Z_n+\ell^2\\
  \frac{Z_n-1-Y_n}{1-p}&=&Z_{n-1}+(\ell-1)Y_{n-1}-(\ell-1)^2
\end{eqnarray*}
Using the first equation, we exclude $Y$'s from the second which turns into a
recurrence
\begin{equation}
  \label{ZZn}
  Z_{n-1} - 2gZ_n + Z_{n+1}+\ell\,\frac{p+(2-p)\ell}{1-p}=0
\end{equation}
whose general solution reads
\begin{equation}
  \label{Zn-sol}
  Z_n=\frac{\ell}{p} + Z_+\lambda_+^n + Z_-\lambda_-^n
\end{equation}

The symmetry $x\leftrightarrow -x$ leads to $Y_0=0$, or
\begin{equation}
  \label{Z1}
  Z_0=(1-p)Z_1+\ell^2
\end{equation}
The periodicity in the large $x$ limit implies that $Z_n$ remains finite for
large $n$. The exponentially growing part of the solution should therefore
vanish, $Z_+=0$. Thus $Z_0=Z_- +\ell/p$ and $Z_1=Z_-\lambda_- +\ell/p$. By
inserting these relations into (\ref{Z1}) we obtain
\begin{equation}
  \label{Z-}
  Z_-=\frac{\ell(\ell-1)}{1-(1-p)\lambda_-}
\end{equation}
Thus the average time in the original problem is given by $\langle
t\rangle=T(0)=Z_0$, or
\begin{equation}
  \label{tav-SBB}
  \langle t\rangle=\frac{\ell}{p}+\frac{\ell(\ell-1)}{1-(1-p)\lambda_-}
\end{equation}
In the limiting cases $p=1$ and $\ell=1$ we indeed recover $\langle
t\rangle=\ell^2$ and $\langle t\rangle=p^{-1}$ [Eqs.~(\ref{TL}) and
(\ref{avt})], respectively.
Finally, the velocity is
\begin{equation}
  \label{v-SBB}
  v=\frac{\langle x \rangle}{\langle t\rangle} =
  \frac{p}{p+(2-p)\ell}\,
  \frac{1-(1-p)\lambda_-+(2-p)\ell}{1-(1-p)\lambda_-+p(\ell-1)}
\end{equation}
Using (\ref{lambda}), one can transform (\ref{v-SBB}) into (\ref{v-main}).

In the initial state all bridges are intact, and a fraction of them remains
intact as the walker moves along.  The walker passes on average $\langle
x\rangle/\ell$ bridges per one burnt bridge. Hence the fraction $I$ of
bridges which forever remains intact approaches
\begin{equation}
   \label{intactexp}
   I=\frac{\langle x\rangle/\ell-1}{\langle x\rangle/\ell}=1-\frac{\ell}{\langle x\rangle}
\end{equation}
in the long time limit.  Using Eq.~(\ref{xav-SBB}) one recasts
(\ref{intactexp}) into Eq.~(\ref{intactic}).

\begin{figure}[htb]
  \includegraphics[width=0.9\linewidth]{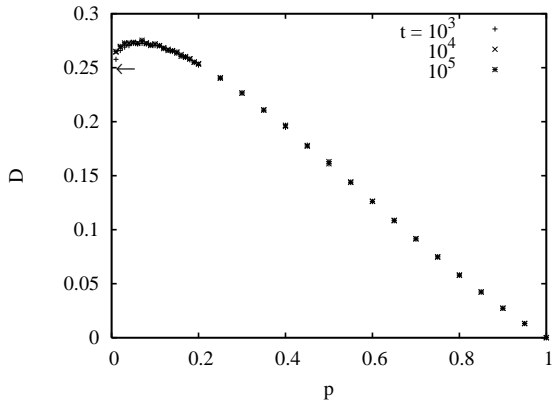}
  \caption{Numerical results for the diffusion coefficient for a system full
    of bridges $(c=1)$ as function of $p$ for several times. The arrow points
    to the analytic result $D(p\to 0)=1/4$. }
  \label{diffp}
\end{figure}

The calculation of the diffusion coefficient seems to be very
challenging. For a system full of bridges ($c=1$), however, the walk is
somewhat analogous to the $p=1$ (and $c<1$) case, and $D$ might be possible
to derive using the approach presented in Sec.~\ref{vD}. The complete
analysis appears to be very cumbersome, but if $p\to 0$ in addition to $c=1$, the
successive burnt bridges are (on average) separated by large gaps and
therefore one can employ a continuous treatment. Following the steps described
in Sec.~\ref{vD} we obtained $D=1/4$. This predictions agrees with
simulations. Interestingly (see Fig.~\ref{diffp}),
$D$ is a non-monotonous function of $p$.

Solving the stochastic burnt bridge model for randomly distributed bridges
does not look possible in the realm of the above framework. Indeed, instead of
working with ordinary {\em deterministic} recurrences like (\ref{ABn}) one
has to tackle {\em stochastic} recurrences (see Appendix \ref{stoch}).

\section{Discussion}
\label{disco}

Our current understanding of the stochastic burnt bridge model is certainly
incomplete --- only the periodic case is somewhat tractable, albeit even in this
situation we do not know how to compute various interesting quantities like
the diffusion coefficient or the probability that in the final state two
nearest burnt bridges are separated by $k$ intact bridges.

In many biological applications, molecular motors move along a homogeneous
polymer filament (kinesin and myosin are classical examples \cite{motors}),
while in other applications the track is inhomogeneous (this particularly
happens when motors move along DNA). It would be interesting to study the
burnt bridge model when in addition to the disorder related to location of
the bridges there is the disorder associated with hopping rates. Earlier work
on random walkers under the influence of a random force (\cite{D,B} and
references therein) and recent work motivated by single-molecule experiments
on motors moving along a disordered track \cite{harms,KLN} may be useful in that
regard.

\section{acknowledgment}

We are thankful to M.~Stojanovic for bringing Ref.~\cite{sci} to our
attention, and for useful remarks. TA gratefully acknowledges financial support from the Swiss
National Science Foundation under the fellowship 8220-067591.

\appendix

\section{Calculation of $[ t ]_\ell$ and $[ t^2 ]_\ell$ for an interval}
\label{moments}

Let $t$ be the first passage time, namely, the time it takes for the simple
random walk in an interval $[0,\ell]$ starting at site $0$ to reach site
$\ell$ for the first time. Here we compute the first two moments, $[t]_\ell$
and $[t^2]_\ell$, of this random variable. We will present an elementary
approach that does not require the calculation of the complete first passage
probability \cite{redner}.  (The calculations in Sec.~\ref{SBB}
can also be considered as a generalization of this method.)

The process can be understood in terms of a random variable $t(x)$, which is
the time it takes to reach site $\ell$ if the walker starts at site $x$. As the
walker from site $x$ steps equally probably to either side
\begin{equation}
  \label{expli}
  t(x) =
  \begin{cases}
    t(x-1)+1  & {\rm probability~ 1/2}\\
    t(x+1)+1 & {\rm probability~ 1/2} ~,
  \end{cases}
\end{equation}
and for the average time $T(x)= [t(x)]_\ell$ we arrive at the recursion
formula
\begin{equation}
  \label{master}
  T(x)=\frac{1}{2}\left[T(x-1)+T(x+1)\right]+1~.
\end{equation}
This master equation holds for $1\leq x\leq \ell-1$ while for $x=0$ it should
be replaced by
\begin{equation}
  \label{master1}
  T(0)=T(1)+1
\end{equation}
since when the walker is at site 0, it always makes the step to the right.  The
recurrence (\ref{master})--(\ref{master1}) is supplemented by the boundary
condition $T(\ell)=0$.

Equation (\ref{master1}) can be re-written in the general form (\ref{master})
if $T(-1)=T(1)$. Further one finds that Eq.~(\ref{master}) holds for $x=-1$
if $T(-2)=T(2)$, and generally Eq.~(\ref{master}) applies for all $|x|\leq
\ell-1$. Hence we seek a solution invariant under the transformation
$x\leftrightarrow -x$; the absorbing boundary conditions are $T(\pm\ell)=0$.
(Numerous examples of analyzing equations like (\ref{master}) with absorbing
boundary conditions are described in \cite{redner}.) The solution is very
neat
\begin{equation}
  \label{Tx-sol}
  T(x)=(\ell+x)(\ell-x)
\end{equation}
and in particular
\begin{equation}
  \label{TL}
  [t]_\ell \equiv T(0)=\ell^2
\end{equation}

For the derivation of the second moment, it is again convenient to consider
$T_2(x)=[ t^2(x) ]_\ell$ which is the average square time to reach site
$\ell$ for the first time if the walker starts at position $x$. From
Eq.~(\ref{expli}) one obtains the governing equation for $1\leq x\leq \ell-1$
\begin{equation}
  \label{x}
  0=\frac{1}{2}\,D^2 T_2(x)+T(x-1)+T(x+1)+1~,
\end{equation}
where $D^2F(x)=F(x-1)-2F(x)+F(x+1)$ is the shorthand notation for the
discrete derivative of the second order. For $x=0$ we have
\begin{equation}
  T_2(0) = T_2(1) + 2T(1) + 1 ~.
\end{equation}

We can again seek a solution to Eq.~(\ref{x}) satisfying the symmetry
requirement $x\leftrightarrow -x$ and the absorbing boundary conditions are
$T_2(\pm\ell)=0$.  Using Eq.~(\ref{Tx-sol}) we recast Eq.~(\ref{x}) into
\begin{equation}
  \label{TxM}
  D^2 T_2(x)=4x(x+1)-4x+2-4\ell^2 ~,
\end{equation}
which yields to
\begin{eqnarray*}
  T_2(x)=\frac{1}{3}\,x^2\left[x^2+2-6\ell^2\right]+ [t^2]_\ell
\end{eqnarray*}
with
\begin{equation}
  \label{TL-per}
  [t^2]_\ell =\frac{1}{3}\,\ell^2(5\ell^2-2)  ~.
\end{equation}

The derivation of the moments for the modified burnt bridge model, where the
hopping rule differs from the original model only from site 0, follows the
same lines.  The new rule affects only Eq.~(\ref{master1}) which now becomes
$T(0)=\frac{1}{2}[T(0)+T(1)]+1$. This equation can be recast in the general
form (\ref{master}) if $T(-1)=T(0)$, and overall the symmetry $T(x)=T(-x-1)$
allows us to reduce the problem to solving (\ref{master}) subject to
$T(\ell)=0$ and $T(-\ell-1)=0$.  The solution $T(x)=(\ell+1+x)(\ell-x)$
yields $[t]_\ell=T(0)=\ell(\ell +1)$.

For the second moment the governing equation is given by Eq.~(\ref{x}) for
$1\leq x\leq \ell-1$, and for $x=0$ it is
\begin{equation}
  \label{1}
  T_2(0)=\frac{T_2(0)+T_2(1)}{2}+T(0)+T(1)+1~.
\end{equation}
The boundary condition is $T_2(\ell)=0$.

A solution of Eq.~(\ref{TxM}) invariant under the transformation
$x\leftrightarrow -x-1$ and satisfying the absorbing boundary conditions
$T_2(\ell)=T_2(-\ell-1)=0$ is
\begin{equation}
  \label{Tx}
  D^2 T_2(x)=4x(x+1)+2-4(\ell^2+\ell)
\end{equation}
which is solved to yield
\begin{eqnarray*}
  T_2(x)&=&\frac{1}{3}\,(x-1)x(x+1)(x+2)\\
  &+&[1-2(\ell^2+\ell)]\,x(x+1)+  [t^2]_\ell
\end{eqnarray*}
with
\begin{equation}
  \label{TL-periodic}
  [t^2]_\ell= \frac{1}{3}\,\ell(\ell+1)[5\ell(\ell+1)-1]
\end{equation}

\section{Direct calculation of $v(1,p)$}
\label{direct}

Here we present an alternative, direct calculation of the velocity for a
lattice fully covered with bridges ($c=1$).  At each time
step, the walker makes a move, so after $t$ time steps all
bridges remain intact with probability $(1-p)^t$. Hence the first burning
event would happen at time $(t+1)$ with probability
\begin{equation}
  \label{Bt}
  B(t) = p(1-p)^t
\end{equation}
and thus the average time till the first burning event is
\begin{equation}
  \label{avt}
  [t]=\sum_{t\geq 0} (t+1)p(1-p)^t=p^{-1} 
\end{equation}

We also need the probability distribution $P(x,t)$ of the position of the
walker. As described earlier, we can consider the unconstrained random walk on
the infinite line, and then ``fold'' it at the origin to give
\begin{equation}
  P(x,t) = \left\{
    \begin{tabular}{ll}
      $P_0(x,t) + P_0(-x,t)$~~~~ & for $x>0$\\
      $P_0(0,t)$ & for $x=0$
      \label{fold}
    \end{tabular}
  \right.
\end{equation}
with $P_0(x,t)$ being the probability distribution of the unconstrained
walker. When the walker starts from the origin at time $t=0$, this
probability is
\begin{equation}
  P_0(x,t) = \left\{
    \begin{tabular}{ll}
      $2^{-t} {\displaystyle \binom{t}{ \frac{t+x}{2}}}$~~~~ & for $t+x$ even\\
      $0$ & for $t+x$ odd  
    \end{tabular}
  \right.
\end{equation}
The probability that a bridge burns at time $(t+1)$ when the walker hops
from site $x$ is $P(x,t)B(t)$, and the total probability is obtained after
summing over all $t$:
\begin{equation}
  \label{Bx}
  \mathcal{B}(x) = \sum_{t=0}^\infty P(x,t) B(t) ~.
\end{equation}

If the walker is hopping to the right when the burning occurs, the move is
completed; otherwise the walker remains in its position. Both of these
alternatives occur equiprobably when $x>0$, while when $x=0$ the walker
surely hops to the right. The average final position of the walker is
therefore
\begin{equation}
  \label{avx-gen}
  \langle x\rangle=\mathcal{B}(0) 
  + \sum_{x=1}^\infty \left(x+\frac{1}{2}\right) \mathcal{B}(x) 
\end{equation}
Using (\ref{fold}), (\ref{Bx}), and the identity $\sum_{x\geq
  0}\mathcal{B}(x)=1$ we transform (\ref{avx-gen}) into
\begin{equation*}
  \langle x\rangle= \frac{1}{2}+ \frac{1}{2}\sum_{t=0}^\infty P_0(0,t) B(t) 
  + \sum_{x=-\infty}^\infty |x| \sum_{t=0}^\infty P_0(x,t) B(t) 
\end{equation*}
The first sum reduces to
\begin{equation}
  \label{first}
  \frac{p}{2} \sum_{k=0}^\infty a^{2k} \binom{2k}{k} = \frac{p}{2\sqrt{1-4a^2}} ~,
\end{equation}
where $a=(1-p)/2$. Next we re-write the second sum as
\begin{equation*}
  \sum_{t=0}^\infty B(t) V(t) ~~, ~~~ V(t) = \sum_{x=-t}^t |x| P_0(x,t) 
\end{equation*}
and simplify $V(t)$ by separately considering even and odd times:
\begin{equation*}
  V(t) = \left\{
    \begin{tabular}{ll}
      $2^{-2k+1} \sum\limits_{m=-k}^k |m|~ \binom{2k}{k+m}$~~~  & for $t=2k$\\
      $2^{-2k} \sum\limits_{m=-k}^{k+1} |m|~ \binom{2k+1}{k+m}$ & for $t=2k+1$ 
    \end{tabular}
  \right.
\end{equation*}
Evaluating the sums we obtain
\begin{equation*}
  V(t) = \left\{
    \begin{tabular}{ll}
      $2^{-2k+1} (k+1) {\displaystyle \binom{2k}{k+1}}$~~~ & for $t=2k$\\
      $2^{-2k}  (k+1) {\displaystyle \binom{2k+1}{k+1}}$~~~ & for $t=2k+1$ 
    \end{tabular}
  \right.
\end{equation*}
Putting this into $\sum_{t\geq 0} B(t) V(t)$ we find that the sum is equal to
\begin{eqnarray}
  \label{second}
  &2p& \sum_{k=0}^\infty 
  \left[ a^{2k} (k+1) \binom{2k}{k+1} + a^{2k+1} (k+1) \binom{2k+1}{k+1} \right]\nonumber\\
  &=& 2p \left[ \frac{2a^2 + a}{(1-4a^2)^{3/2}} \right]
\end{eqnarray}
Combining (\ref{first}) and (\ref{second}) we obtain the average displacement
\begin{equation}
  \label{avx-exact}
  \langle x\rangle= \frac{1}{2} + \frac{1}{2}\sqrt{\frac{2-p}{p}} 
\end{equation}
and therefore $v=\langle x\rangle/[t]=p\langle x\rangle$ is indeed given by
(\ref{vp-eq}).

\section{The stochastic burnt bridge model in the case of randomly positioned bridges}
\label{stoch}

The formalism detailed in Sect.~\ref{SBB} for the periodic location of
bridges to the situation formally applies to the situation when bridges are
arbitrarily distributed. Let $(\ell_1-1,\ell_1),
(\ell_1+\ell_2-1,\ell_1+\ell_2)$, etc. be bridge locations.  Away from
bridges the governing equation (\ref{Lx}) is valid while on the boundaries
the modified equations are almost identical to (\ref{n-1}) and (\ref{n}), the
only exception is that $n\ell$ should be replaced by $L_n\equiv
\ell_1+\ldots+\ell_n$.  The average position of the walker $L(x)$ is again a
linear function of $x$ on each interval between neighboring bridges; for
$L_n\leq x\leq L_n+\ell_{n+1}-1$
\begin{equation}
  \label{Lx-linear}
  L(x) = A_n +(x-L_n)B_n 
\end{equation}
The analogs of Eqs.~(\ref{An})--(\ref{Bn}) are
\begin{eqnarray*}
  &&B_{n-1}=[-A_{n-1}+(1-p)A_n+pL_n]/\ell_n\\
  &&A_n=B_n+(1-p)[A_{n-1}+(\ell_n-1)B_{n-1}]+pL_n
\end{eqnarray*}
Using the first equation we exclude $B$'s from the second and thereby recast
it into a recurrence
\begin{eqnarray}
  \label{rec-stoch}
  \frac{A_{n+1}}{\ell_{n+1}}+\frac{A_{n-1}}{\ell_n}&=&
  \frac{A_n}{1-p}\left[p(2-p)+\frac{1}{\ell_{n+1}}+\frac{(1-p)^2}{\ell_n}\right]\nonumber\\
  &+&p+pL_n\left[2-p+\frac{1}{\ell_{n+1}}-\frac{1-p}{\ell_n}\right]
\end{eqnarray}

In the interesting case when $\ell$'s are independent identically distributed
random variables, one must solve the stochastic inhomogeneous recurrence
(\ref{rec-stoch}). Even a homogeneous version of Eq.~(\ref{rec-stoch}) is
analytically intractable.  The additional challenging feature of the
inhomogeneous recurrence (\ref{rec-stoch}) is infinite memory manifested by
factor $L_n$; as a result, it is not clear how to find a particular solution
of Eq.~(\ref{rec-stoch}) which is required if one wants to reduce the problem
to solving a homogeneous version of Eq.~(\ref{rec-stoch}).

The case of weak disorder is probably exceptional, e.g., it should be
possible to compute the growth rates $\lambda_\pm$. One can, however, avoid
such a lengthy analysis by noting that in the present context the condition
of weak disorder implies that bridges are located almost periodically and
their concentration is small ($c\ll 1$). Assuming additionally that the
bridge burning probability is not anomalously small, so that $c\ll p$, an
argument presented after Eq.~(\ref{v-asymp}) shows that in the leading order
the burnt bridge model must be recovered. Thus $v\approx c$ and $D\approx (1-c^2)/3$.


\begin{thebibliography}{99}
  
\bibitem{saw} 
  N. Madras and G. Slade, {\it The Self-Avoiding Walk}
  (Birkhauser, Boston, 1992).
    
\bibitem{mai} 
  J.~Mai, I.~M.~Sokolov, and A.~Blumen, Phys. Rev. E {\bf 64},
  011102 (2001).

\bibitem{fox} 
  R.~Fox, Phys. Rev. E {\bf 57}, 2177--2203 (1998).

\bibitem{siam} 
  T.~C.~Elston and C.~S.~Peskin, SIAM J. Appl. Math. {\bf 60},
  842--867 (2000); T.~C.~Elston, D.~You, and C.~S.~Peskin, SIAM J. Appl.
  Math.  {\bf 61}, 776--791 (2000).

\bibitem{der}
  I. Der\'enyi and T. Vicsek, Physica A {\bf 249}, 397 (1998); 
  Proc. Natl. Acad. Sci. USA {\bf 93}, 6775 (1996). 
  
\bibitem{lipo}  
  S. Klumpp, T. M. Nieuwenhuizen, and R. Lipowsky, cond-mat/0502527.

\bibitem{motors}
  J.~Howard, {\it Mechanics of Motor Proteins and the Cytoskeleton}
  (Sinauer Associates, Sunderland, MA, 2001).

\bibitem{sci} 
  S.~Saffarian, I.~E.~Collier, B.~L.~Marmer, E.~L.~Elson, and
  G.~Goldberg, Science {\bf 306}, 108--111 (2004).
  
\bibitem{again} 
  The continuum approximation again manages to get one answer
  right --- now it is exact in the random case and approximate in the
  periodic case.

\bibitem{ctrw} 
  H.~Scher and M.~Lax, Phys.\ Rev.\ {\bf 137}, 4491 (1973); 
  E.~W.~Montroll and G.~H.~Weiss, J. Math.\ Phys. {\bf 6}, 167 (1965).

\bibitem{Hughes} 
  B. D. Hughes, \newblock {\it Random Walks and Random
    Environments, Vol. 1: Random Walks} (Clarendon Press, Oxford, 1996).

\bibitem{sci2} More precisely, $G(t)-1\propto C(t)$, where $G(t)$ is the
  correlation function from Ref.~\cite{sci}.

\bibitem{D} 
  B.~Derrida, J. Stat.\ Phys. {\bf 31}, 433 (1983).

\bibitem{B} 
  J.~P.~Bouchaud, A.~Comtet, A.~Georges, and P.~Le~Doussal, Ann.\ Phys. 
  {\bf 201}, 285 (1990).
  
\bibitem{harms} 
  T. Harms and R. Lipowsky, Phys.\ Rev.\ Lett.\ {\bf 79}, 2895 (1997).

\bibitem{KLN} 
  Y.~Kafri, D.~K.~Lubensky, and D.~R.~Nelson, Biophys. J. {\bf 86},
  3373 (2004);  Phys. Rev. E {\bf 71}, 041906 (2005).

\bibitem{redner} 
  S.~Redner, {\it A Guide to First-Passage Processes}
  (Cambridge University Press, New York, 2001).

\end{thebibliography}
\end{document}